\documentclass[twoside]{dis07}
\usepackage[latin1]{inputenc}
\usepackage[dvips]{graphicx,epsfig,color}
\usepackage{wrapfig,rotating}
\usepackage{amssymb,amsmath,array}
\usepackage{xspace}
\usepackage{subfigure}

\newcommand{\lam}{\ensuremath{\Lambda} \xspace}

\newcommand{\alam}{\ensuremath{\overline{\Lambda}}\xspace}

\newcommand{\pion}{\ensuremath{\pi}\xspace}

\newcommand{\kz}{\ensuremath{\mathrm{K^{0}_{S}}}\xspace}

\newcommand{\midrap}{\ensuremath{{|{\bf y}|<0.5}}\xspace}
\newcommand{\pt}{\ensuremath{p_{T}}\xspace}
\newcommand{\mt}{\ensuremath{m_{T}}\xspace}

\newcommand{\pp}{\ensuremath{p+p}\xspace}

\newcommand{\ee}{\ensuremath{e^{+}+e^{-}}\xspace}
\newcommand{\sqsRhic}{\ensuremath{\sqrt{s}=200}\xspace}

\begin{document}
\title{STAR identified particle measurements at high transverse momentum in p+p
$\sqrt{s}=200$ GeV}

%***********************************************************************
% AUTHORS INFORMATION AREA
%***********************************************************************
\author{Mark Heinz for the STAR Collaboration
%
%
% DO NOT MODIFY THE FOLLOWING '\vspace' ARGUMENT
\vspace{.3cm}\\
%
% Addresses and institutions (remove "1- " in case of a single institution)
Yale University, Physics Department, WNSL,\\
272 Whitney Ave, New Haven, CT 06520, USA }
%***********************************************************************
% END OF AUTHORS INFORMATION AREA
%***********************************************************************

\maketitle

\begin{abstract}
We present the STAR measurement of transverse momentum spectra at
mid-rapidity for identified particles in \pp collisions at \sqsRhic
GeV. These high statistics data are ideal for comparing to existing
leading- and next-to-leading order (NLO) perturbative QCD
calculations. Next-to-leading models have been successful in
describing inclusive hadron production using parameterized
fragmentation functions (FF) for quarks and gluons. However, in
order to describe baryon spectra at NLO, knowledge of flavor
separated FF is essential. Such FF have recently been parameterized
using data by the OPAL experiment and allow for the first time to
obtain good agreement between NLO and identified baryons from \pp
collisions.
\end{abstract}

\section{Introduction}\label{intro}

Perturbative QCD has proven to be successful in describing inclusive
hadron production in elementary collisions. Within the theory's
range of applicability, calculations at next-to-leading order (NLO)
have produced accurate predictions for transverse momentum spectra
of inclusive hadrons at different energy scales \cite{Marco:SQM04}.
With the new high statistics proton-proton data at \sqsRhic GeV
collected by STAR, we can now extend the study to identified baryons
and mesons to \pt $\sim$9 GeV/c. Perturbative QCD calculation apply
the factorization ansatz to calculate hadron production and rely on
three ingredients. The first part are the non-perturbative parton
distribution functions (PDF) which are obtained by parameterizations
of deep inelastic scattering data. They describe quantitatively how
the partons share momentum within a nucleus. The second part, which
is perturbatively calculable, consists of the parton cross-section
amplitude evaluated to LO or NLO using Feynman diagrams. The third
part consists of the non-perturbative Fragmentation functions (FF)
obtained from \ee collider data using quark-tagging algorithms.
These parameterized functions are sufficiently well known for
fragmenting light quarks, but less well known for fragmenting gluons
and heavy quarks. Recently, Kniehl, Kramer and P\"otter (KKP) have
shown that FF are universal between \ee and \pp collisions
\cite{KKP:01}. At leading-order, we compare to string fragmentation
models such as PYTHIA to investigate the dependence between hadrons
and underlying parton-parton interactions \cite{Pythia}. In the
string fragmentation approach the production of baryons is
intimately related to di-quark production from strings. They then
combine with a quark to produce a baryon. In NLO calculations, the
accuracy of a given baryon cross-section is based on the knowledge
of that specific baryon fragmentation function (FF) extracted \ee
collisions.

\section{Data Analysis}\label{analysis}

The present data were reconstructed using the STAR detector system
which is described in more detail elsewhere \cite{STAR2}. The main
detectors used in this analysis are the Time Projection Chamber
(TPC) and the Time of Flight detector (TOF). A total of 14 million
non-singly diffractive (NSD) events were triggered with the STAR
beam-beam counters (BBC) requiring two coincident charged tracks at
forward rapidity ($3.3 < |\eta| < 5.0$). By simulation it was
determined that the trigger measured 87\% of the 30.0$\pm$3.5mb NSD
cross-section. The offline primary vertex reconstruction was 76\%
efficient which lead to a total usable event sample of
$7\times10^{6}$ events. Protons and pions in this analysis were
identified using the TOF detector at \pt below 2.5 GeV/c and the TPC
using the relativistic rise dE/dx at higher \pt. Details of both
methods are described in \cite{STAR:tof,STAR:relrise}. At \pt $\sim
3$ GeV/c the pion dE/dx is about 10-20\% higher than that of kaons
and protons due to the relativistic rise, resulting in a few
standard deviations of seperation. The strange particles were
identified from their weak decay to charged daughter particles. The
following decay channels and the corresponding anti-particles were
analyzed: $\mathrm{K^{0}_{S}} \rightarrow \pi^{+} + \pi^{-}$ (b.r.
68.6\%), $\Lambda \rightarrow p + \pi^{-}$(b.r. 63.9\%) . Particle
identification of the daughters was achieved by requiring the
TPC-measured dE/dx to fall within the 3$\sigma$-bands of the
theoretical Bethe-Bloch parameterizations. Further background in the
invariant mass was removed by applying topological cuts to the decay
geometry.

\begin{figure}[h]
\centering \mbox{
\subfigure[]{\includegraphics[width=6cm]{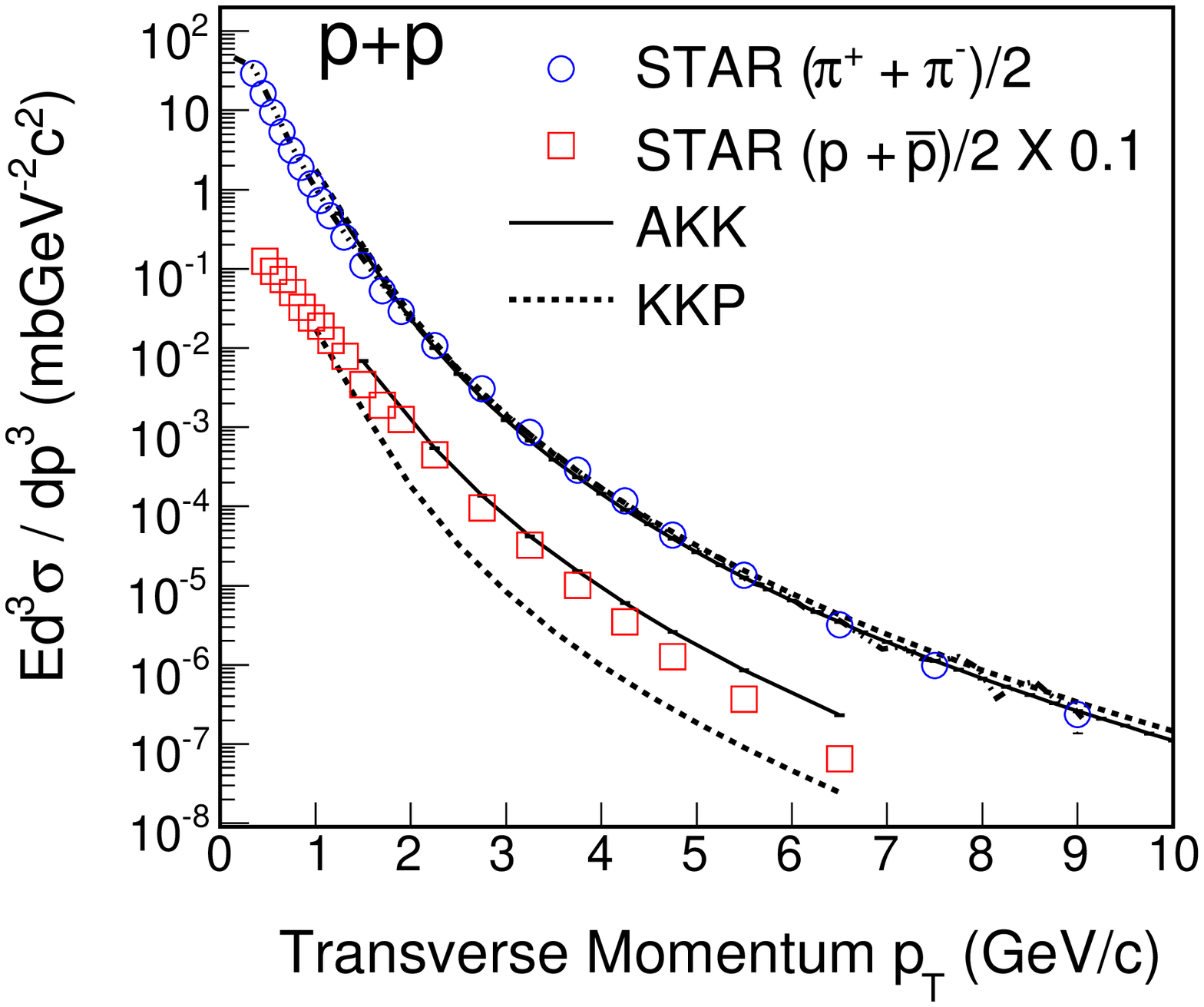}}
\quad
\subfigure[]{\includegraphics[width=6cm]{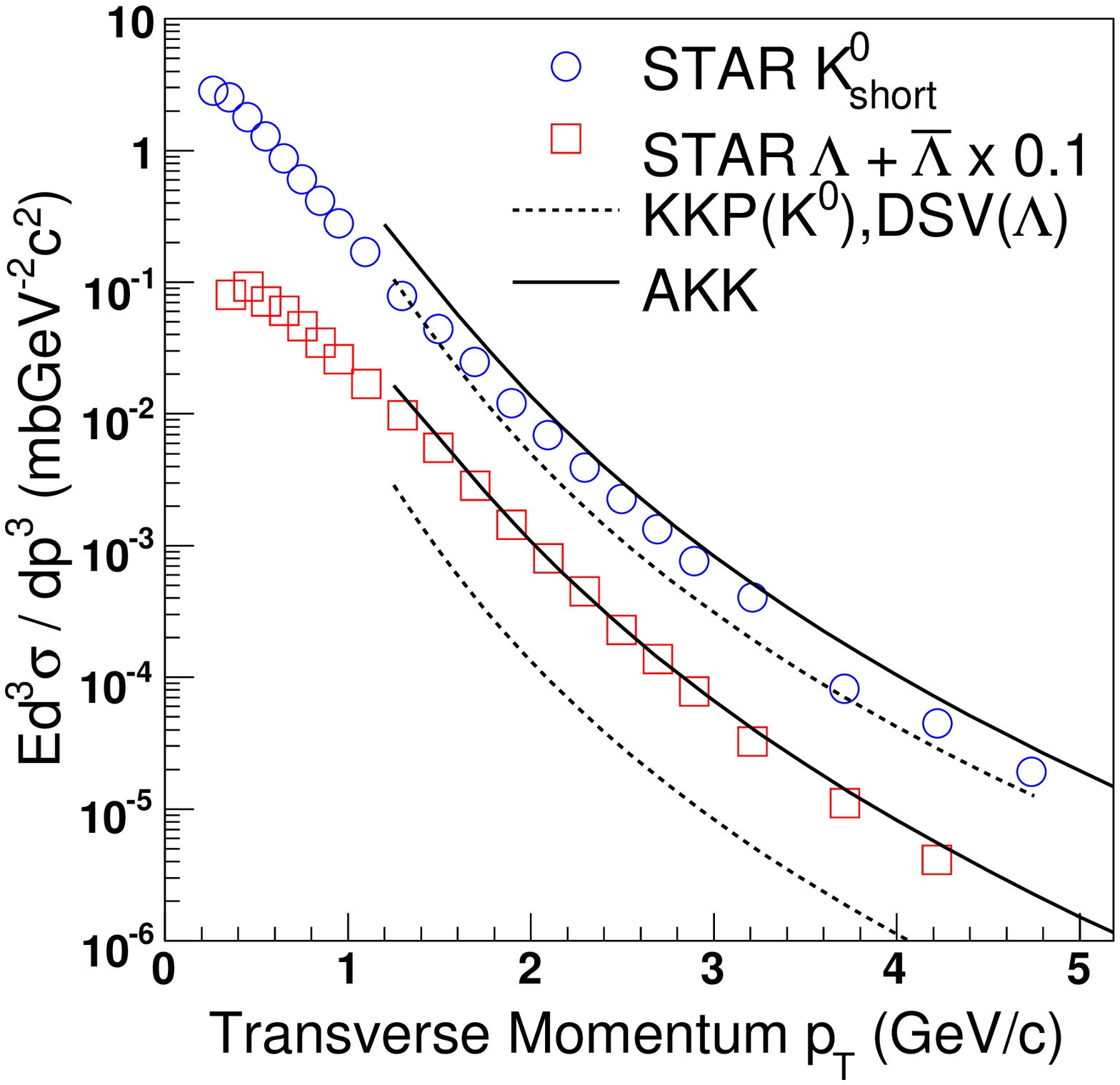}} }
\caption{(a)$(p+\bar{p})$ and $(\pi^{+}+\pi^{-})$ \pt-spectra at
\midrap from NSD p+p at 200 GeV \cite{BedangaPLB:06} (b) \kz and
\lam \pt spectra at \midrap compared to NLO calculations using KKP,
DSV and AKK fragmentation functions. \cite{StrangePRC:07} }
\label{fig:nlo}
\end{figure}

\section{Results}
\subsection{Comparison to next-to-leading order}

In figure \ref{fig:nlo} we compare our transverse momentum spectra
to recent next-to-leading order calculations using two different
fragmentation functions (FF). The previous ones were by
Kniehl-Kramer-Poetter (KKP) for pions, kaons and protons and from
deFlorian-Stratmann-Vogelsang (DSV) for \lam \cite{KKP:00,DSV}. More
recently Albino-Kramer-Kniehl (AKK) \cite{AKK} have published FF
based on the light quark-flavor tagged data from the OPAL
Collaboration \cite{OPAL:00}. Clearly, these newer parameterizations
improve the description of the baryon data considerably. In order to
achieve this agreement with the data, the initial gluon to \lam
fragmentation function is determined by fixing it's shape to that of
the proton, and then varying the normalization for the best fit. A
scaling factor of 3 with respect to the proton is necessary to
achieve agreement with STAR data. However, this modified FF is then
tested by comparing to the $\Lambda$ measurement from $p+\bar{p}$ at
$\sqrt{s} = 630$ GeV and agrees well \cite{AKK}.

\subsection{Baryon to meson ratios vs \pt}

\begin{figure}[h]
\centering \mbox{
\subfigure[]{\includegraphics[width=5.5cm]{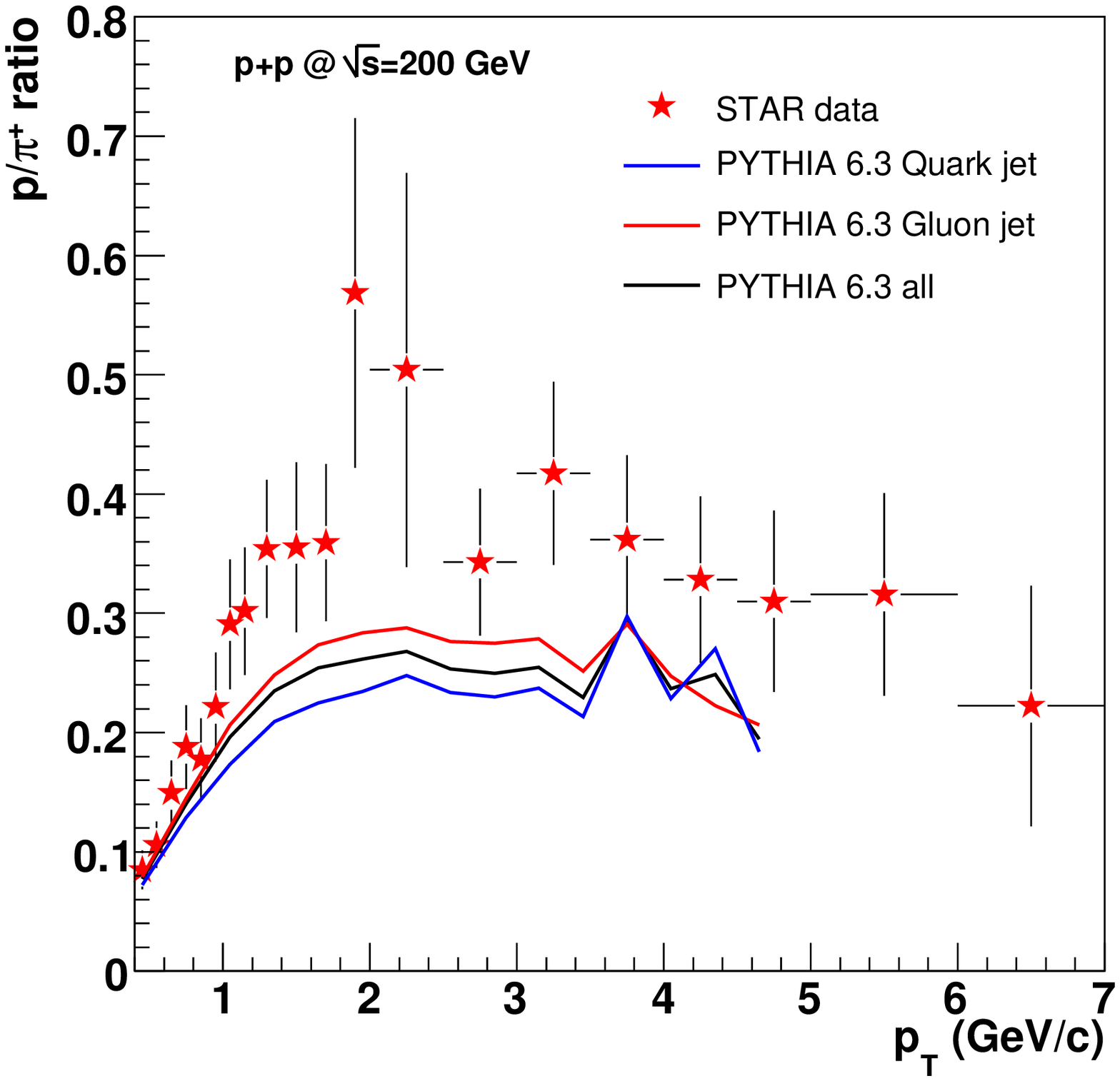}}
\quad
\subfigure[]{\includegraphics[width=5.5cm]{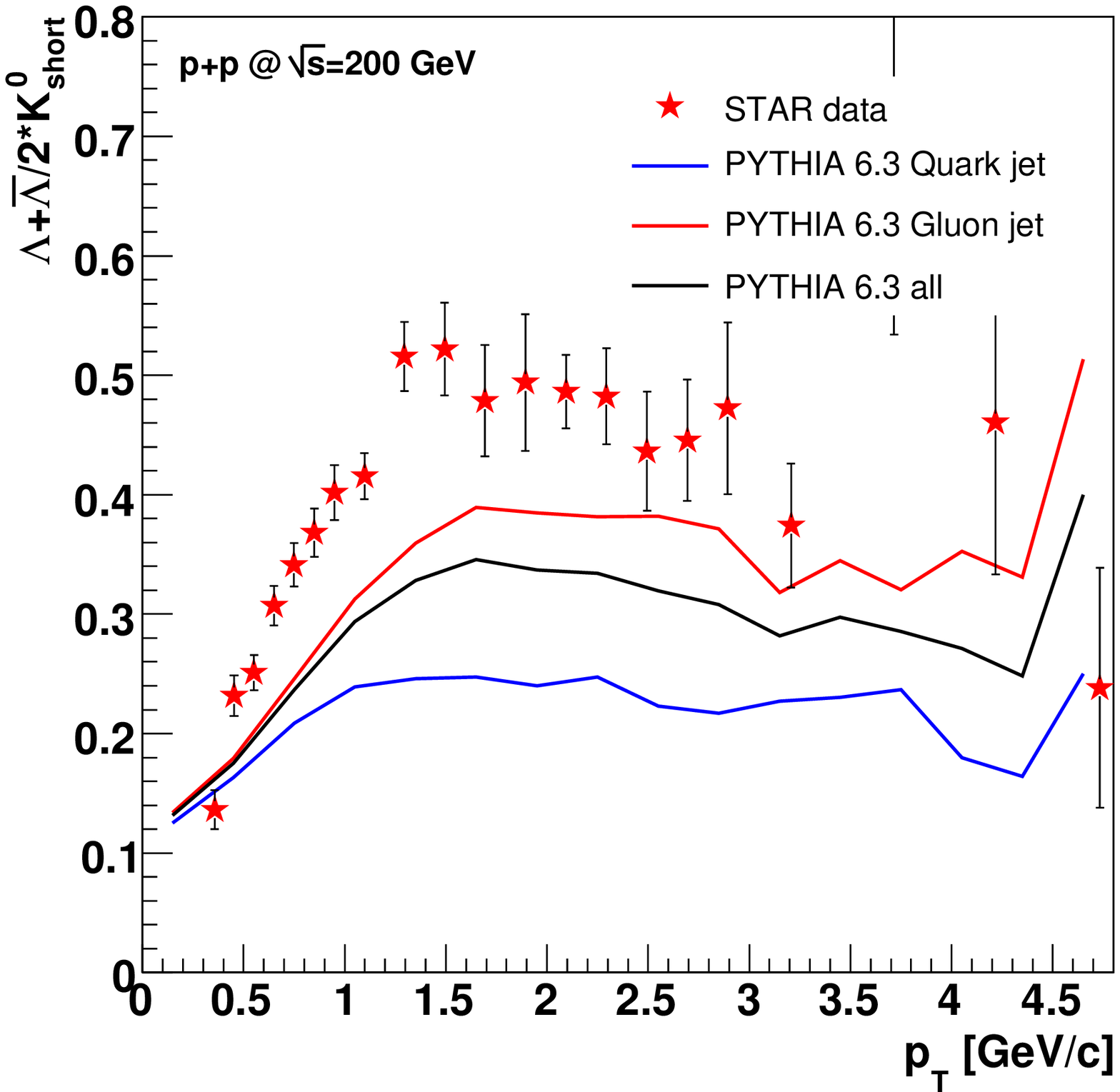}} }
\caption{(a) p/\pion ratio compared to PYTHIA for different event
samples, (b) $(\lam+\alam)/2\cdot\kz$ ratio compared to PYTHIA.}
\label{fig:bmratio}
\end{figure}

In order to further investigate the sensitivity of the baryon
spectra to the fragmentation of gluons, we used a leading-order (LO)
Monte Carlo simulator, PYTHIA. PYTHIA 6.3 generates events by using
$2\rightarrow2$ LO parton processes plus additional leading-log
showers and multiple interactions. We define a ``Gluon-jet" event as
one where the underlying partons are g-g or g-q and a ``Quark-jet"
event one where the underlying partons are q-q. According to the
model default settings the \pp events at our energy are dominated by
gluon-jets (62\%) with respect to quark-jets (38\%). Figure
\ref{fig:bmratio} compares baryon-to-meson ratios to three different
event types from PYTHIA. In both cases the overall ratio in the data
is significantly larger at \pt$\sim$1-3 GeV/c than the PYTHIA
result. In addition, this shows that pure gluon jet events will
produce a larger baryon-to-meson ratio than quark jet events.

\subsection{Transverse mass (\mt) scaling}
\begin{figure}[ht]
\centering \mbox{
\subfigure[]{\epsfig{figure=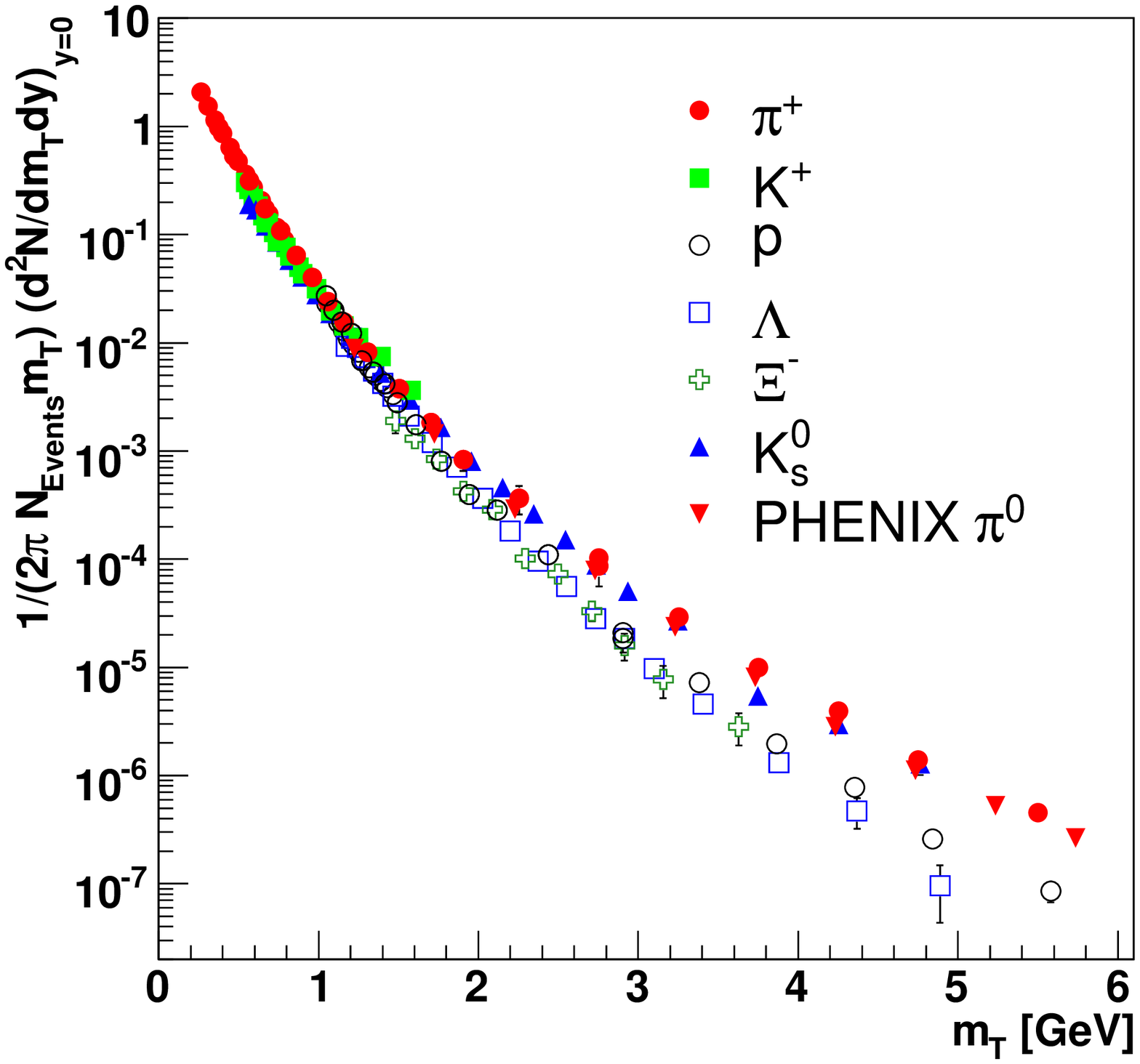,width=4.5cm}}
\subfigure[]{\epsfig{figure=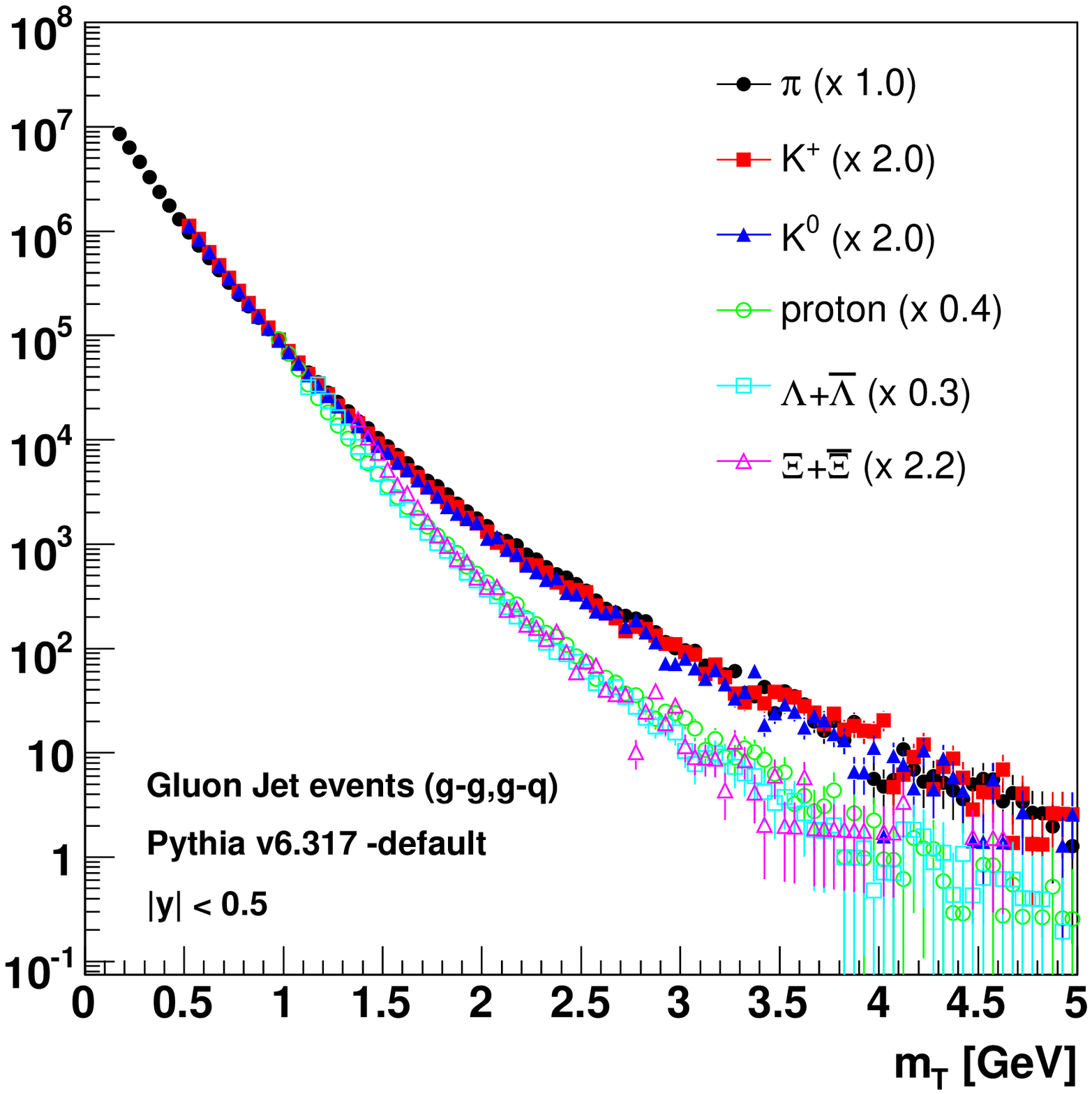, width=4.5cm}}
\subfigure[]{\epsfig{figure=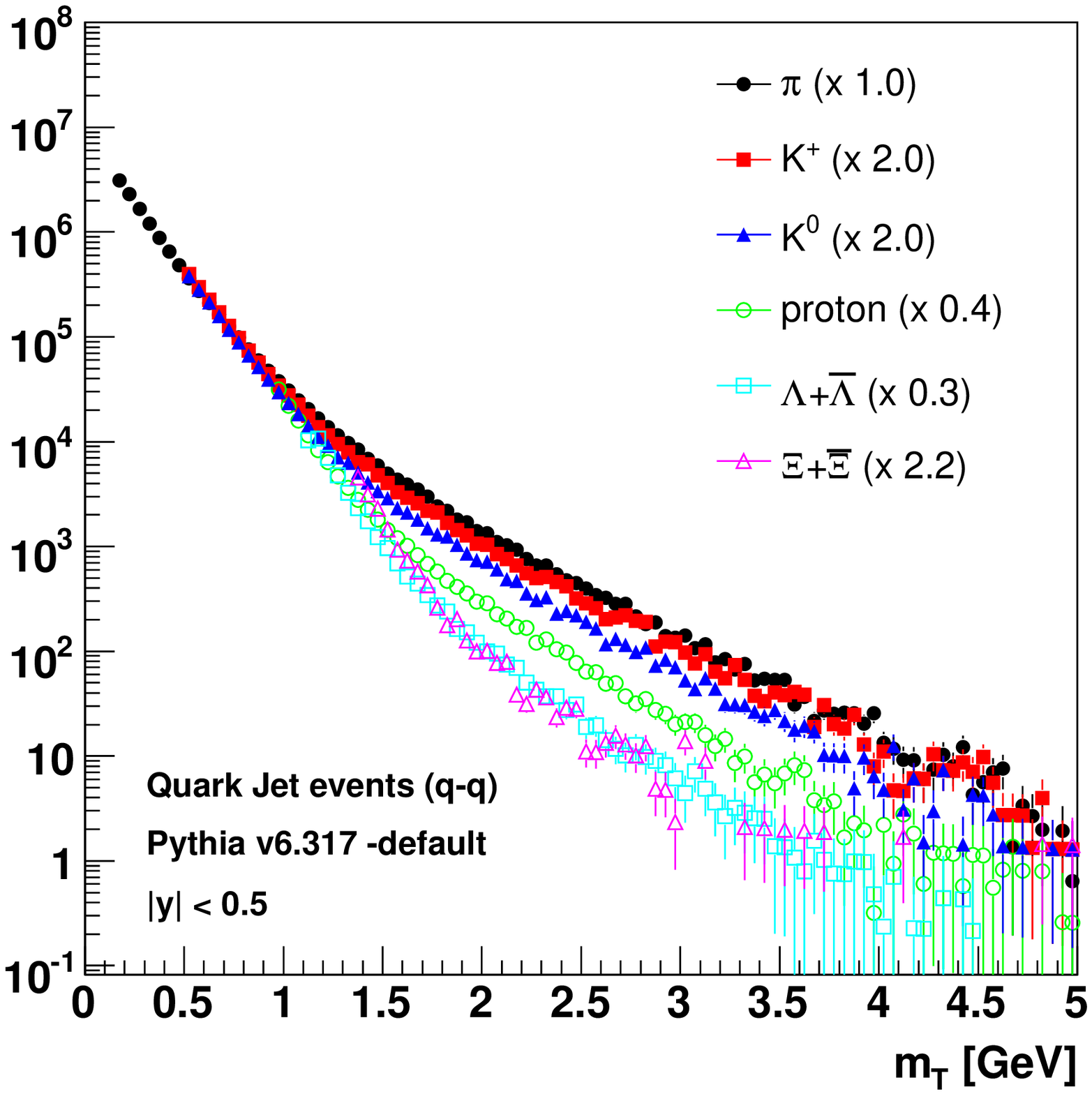, width=4.5cm}}}
\caption{(a)Arbitrarily scaled \mt spectra for baryons and mesons
from $p+p$ collisions at $\sqrt{s} = 200$ GeV. (b) Scaled
\mt~spectra for ``Gluon-jet" events from PYTHIA. (c) Scaled
\mt~spectra for ``Quark-jet" events from PYTHIA.}
\label{fig:mtscaling}
\end{figure}

Universal transverse mass scaling of particle spectra was previously
seen in \pp collisions at lower ISR-energies \cite{Mtscaling}. We
have compiled STAR identified particle spectra to investigate
\mt-scaling. The particle spectra were arbitrarily normalized to
pion spectra at \mt= 1 GeV. Interestingly we observe that a
splitting occurs at $\sim$2 GeV and that the meson spectra are
harder than the baryon spectra. We compared this result to PYTHIA
simulations scaled in the same manner. We again observe that gluon
jets will fragment very differently into baryons and mesons than
quark jets. For gluon jets, there is a clear shape difference
between baryons and mesons consistent with the data. For quark jets,
the shape difference is modified by an additional dependency on the
mass of the produced particle. This may be a further indication that
we observe dominance of gluon jets in \pp collisions at RHIC
energies.

\section{Summary}
We have shown that the theoretical description of identified baryons
and mesons in \pp collisions has recently improved thanks to new NLO
calculations using light quark-flavor tagged fragmentation
functions. Considerable uncertainties remain in the high-z
($p_{hadron}/p_{parton}$) range of the gluon FF for baryons. It
appears that previous baryon-FF extracted from \ee data are
inconsistent with STAR's \pp data, indicating that RHIC is a
valuable test of FF. Arbitrarily scaled \mt~spectra for strange
particles exhibit partial \mt~scaling and confirm the dominance of
gluon jets in \pp and therefore the importance of understanding
gluon fragmentation.

% ****************************************************************************
% BIBLIOGRAPHY AREA
% ****************************************************************************

\begin{footnotesize}
% IF YOU DO NOT USE BIBTEX, USE THE FOLLOWING SAMPLE SCHEME FOR THE REFERENCES
% ----------------------------------------------------------------------------

% ----------------------------------------------------------------------------

\end{footnotesize}

% ****************************************************************************
% END OF BIBLIOGRAPHY AREA
% ****************************************************************************


\begin{thebibliography}{99}
% Please replace the numbers for   contribId   and   sessionId
% in the following URL. You can get this information by going to
% http://indico.cern.ch/confAuthorIndex.py?confId=9499
% and search for your contribution and click on the title
% Be aware: '&amp;' must be replaced by simple '&' as in example below
\bibitem{url} Slides: \\
\verb$http://indico.cern.ch/contributionDisplay.py?contribId=42&sessionId=8&confId=9499$
%------- replace following references ;-)

%--section 1 Intro
\bibitem{Marco:SQM04}M.~van Leeuwen for the STAR Collaboration, {\it J. Phys. G: Nucl. Part. Phys.} \textbf{31} (2005) S881

%universality paper
\bibitem{KKP:01}B.~A.~Kniehl, G.~Kramer and B.~Potter, {\it Nucl. Phys.} B {\bf 597}(2001) 337

\bibitem{Pythia}T.~Sjostrand and P.~Z.~Skands, {\it Eur. Phys. J.} C {\bf 39},
(2005) 129

\bibitem{STAR2}K.H.~Ackermann et al (STAR Collaboration), {\it  Nucl. Instrum. Meth.} A{\bf 499}
(2003) 624

\bibitem{KKP:00}B.~A.~Kniehl, G.~Kramer and B.~Potter, {\it Nucl. Phys.} B {\bf 582}(2000) 514


%----STAR references
\bibitem{STAR:tof}J.~Adams et al. (STAR collaboration), {\it  Phys. Lett.} B {\bf 616},
(2005) 8
\bibitem{STAR:relrise}M.~Shao (STAR collaboration), nucl-ex/0505026

\bibitem{BedangaPLB:06}J.~Adams et al. (STAR collaboration), {\it  Phys. Lett.} B {\bf 637},
(2006) 161
\bibitem{StrangePRC:07}B.~Abelev et al. (STAR collaboration), {\it  Phys. Rev.} C {\bf 75},
(2007) 064901

%----section NLO
\bibitem{DSV}
  D.~de Florian, M.~Stratmann and W.~Vogelsang, {\it Phys. Rev.} D {\bf 57}, (1998) 5811

\bibitem{AKK}S.~Albino et al., {\it Nucl. Phys.} B{\bf 734}, (2006) 50

\bibitem{OPAL:00}G.~Abbiendi {\it et al.}  [OPAL Collaboration], {\it Eur. Phys. J.} C {\bf 16}, (2000) 407

%-----Mt scaling
\bibitem{Mtscaling}
  G.~Gatof and C.Y.~Wong {\it Phys. Rev.} D {\bf 46}, (1992) 997

\end{thebibliography}
\end{document}